\documentclass[prl,aps,floats,twocolumn]{revtex4}

\usepackage[dvips]{graphicx}
\usepackage{amssymb} 
\usepackage{amsmath}
\usepackage{caption}
\usepackage{ctable}

\begin{document}

\title{On Interstellar Quantum Communication and the Fermi Paradox}

\author{Latham Boyle}

\affiliation{Higgs Centre for Theoretical Physics, University of Edinburgh, Edinburgh, UK}
\affiliation{Perimeter Institute for Theoretical Physics, Waterloo, Ontario, Canada}

\date{August 2024}
                            
\begin{abstract}
  Since it began \cite{CocconiMorrison}, the search for extraterrestrial intelligence (SETI) has focused on interstellar {\it classical} communication.  Recently, Berera \cite{Berera:2020rpl} pointed out that, at certain frequencies, photon qubits can retain their quantum coherence over interstellar (and even intergalactic) distances, raising the prospect of interstellar {\it quantum} communication.  This is an intriguing possibility, since quantum communication permits certain tasks that would be impossible with classical communication, and allow exponential speed-ups for others.  (We suggest some motivations in the interstellar context.)  But quantum coherence alone is not sufficient for quantum communication: here, for the first time, we analyze the {\it quantum capacity} $Q$ of an interstellar channel.  We point out that, to have non-zero quantum capacity $Q>0$, interstellar communication over a distance $L$ must use wavelengths $\lambda < 26.5\,cm$ (to avoid depolarization by the cosmic microwave background), and {\it enormous} telescopes of effective diameter $D>0.78\sqrt{\lambda L}$ (to satisfy quantum erasure constraints).  For example, for two telescopes of diameter $D$ on Earth and Proxima Centauri, this implies $D>100\,km$!  This is a technological threshold that remains to be crossed in order for reliable one-way quantum communication to become possible, and suggests a fundamental new resolution of the Fermi paradox.
\end{abstract}
\maketitle

\maketitle

\section{Introduction}

In 1948 Shannon invented the modern theory of {\it classical} information \cite{Shannon}, and in 1959 Cocconi\&Morrison initiated the search for extraterrestrial intelligence (SETI) by noting that existing human technology (radio transmissions \cite{CocconiMorrison} or laser signals \cite{SchwartzTownes}) could be used to send or receive interstellar classical communications.  Over the subsequent decades, physicists have realized that classical information theory is just a part of a much richer, still nascent subject -- {\it quantum} information theory \cite{Nielsen:2012yss} -- and it is natural to ask whether it is also possible to send or receive interstellar {\it quantum} communications.  A first step in this direction was taken by Berera \cite{Berera:2020rpl} (see also \cite{Berera:2021xqa, Berera:2022nzs, Berera:2023fhd, Hippke:2021igp}), who pointed out that -- in certain frequency ranges -- photon qubits could be transmitted over interstellar (or even intergalactic) distances without losing their quantum coherence, and suggested the possibility of searching for such interstellar quantum communications.

The possibility of interstellar quantum communication is intriguing because it expands the notion of interstellar communication in fundamental ways.  First, it is already known to permit many tasks that are impossible with classical communication alone, including quantum cryptography \cite{Bennett:2014rmv, Ekert:1991zz}, quantum teleportation \cite{Bennett:1992tv}, superdense coding \cite{Bennett:1992zzb}, remote state preparation \cite{Bennett:2001bbn}, entanglement distillation/purification \cite{Bennett:1995tk, Bennett:1995ra, Bennett:1996gf}, or direct transmission of (potentially highly complex, highly entangled) quantum states ({\it e.g.}\ the results of complex quantum computations).  Second, protocols based on quantum communication are exponentially faster than those based on classical communication for some problems/tasks \cite{Raz99}, in particular as measured by the one-way classical communication complexity \cite{Bar-Yossef, Gavinsky, Regev} (the number of bits that must be transmitted one-way, from sender to receiver, to solve a problem or carry out a task -- possibly the notion most pertinent to interstellar communication).

The ability of a quantum communication channel to transmit quantum information is determined by its quantum capacity $Q$.  Using constraints on {\it quantum erasure channels} \cite{Bennett:1997mm} and the known properties of the interstellar medium \cite{Ryter}, we show that, in order for an interstellar communication channel to have non-vanishing quantum capacity ($Q>0$), the exchanged photons must lie within certain allowed frequency bands (see Fig. 1), and the effective diameter $D$ of the exchanging telescopes must be enormous: $D>0.78\sqrt{\lambda L}$, where $\lambda$ is the photon wavelength and $L$ is the distance from sender to receiver.  (For a ground-based telescope, and with $L$ the distance to the closest star Proxima Centauri, this requires a diffraction-limited telescope of diameter $D>100km$!)  Using constraints on {\it quantum depolarizing channels} \cite{Bennett:1996gf, Bennett:1995ra, Ekert:1996pg, Knill:1996ny}, and properties of the diffuse astrophysical background radiation \cite{Lequeux}, we show that $Q>0$ requires the exchanged photons to have wavelength $\lambda<26.5~{\rm cm}$ (dominantly due to the cosmic microwave background).  

Thus, our galaxy and universe {\it do} permit interstellar quantum communication, but the above constraints impose a stringent technological threshold we have not yet reached (in particular, we have not yet built a sufficiently large diffraction-limited telescope).  We will see why this suggests a new resolution of the Fermi paradox.

\section{Quantum Capacity of an Interstellar Communication Channel}

The {\it capacity} of a communication channel is the maximum rate at which it can reliably transmit information from sender to receiver (see \cite{Nielsen:2012yss, Bennett:1997mm} for details).  The {\it classical} capacity $C$ is the maximum rate at which classical bits can be communicated.  The {\it quantum} capacity $Q$ is the maximum rate at which qubits can be communicated.
  
In the classical case, two-way communication between sender and receiver does not improve the forward classical capacity $C$.  But in the quantum case, it {\it can}: so $Q$ is the capacity for {\it forward} quantum communication (unassisted by classical communication, or assisted only by {\it forward} classical communication); but one also defines $Q_2$, the quantum capacity of the channel assisted by two-way classical communication, which can be larger than $Q$. (In particular, we can have $Q_2>0$ when $Q=0$.)

Now consider an interstellar quantum communication channel: two telescopes of diameter $D_1$ and $D_2$, separated by a distance $L$, exchanging photons of wavelength $\lambda$.  (Photons can encode quantum states in multiple ways.  Mathematically, the most natural choice would be to use the fact that each photon is intrinsically a two-state system, with positive and negative helicity as the two qubit's basis states $|0\rangle$ and $|1\rangle$, since these are the eigenstates of the ``little group" of symmetries preserving the line of sight between sender and receiver.  But this is just an example to keep in mind for the sake of concreteness -- the following analysis does not rely on this choice.)  In this section, we determine the constraints on such a channel by considering two model quantum channels in turn: the quantum erasure channel and the depolarizing channel.  

\begin{figure}
  \begin{center}
    \includegraphics[width=3.0in]{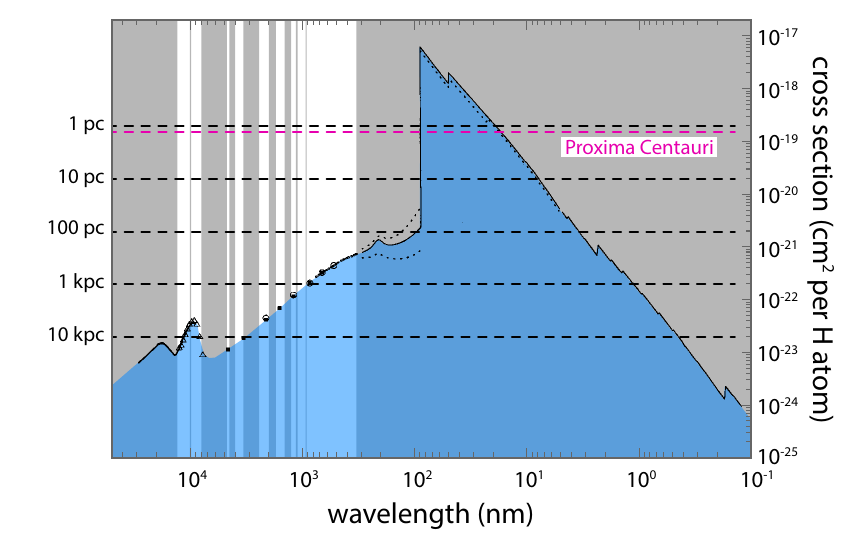}
  \end{center}
  \caption{Quantum communication with $Q>0$, over distance $L$, is impossible at wavelengths where the horizontal line corresponding to $L$ lies within the blue shaded region (summarizing the Milky Way ISM's extinction curve).  Gray regions are off limits from the ground.  Adapted from \cite{Ryter, Lequeux}, with data from \cite{Draine:1984wn, Rieke:1985wv, MartinWhittet, CardelliClaytonMathis, Bastiaansen, FitzpatrickMassa, Rumph, Morrison:1983hg}.}
\label{WindowFig}
\end{figure}

{\bf Quantum erasure channel.}  
This is an idealized channel in which each input qubit $|\psi\rangle=\alpha|0\rangle+\beta |1\rangle$ is, with probability $\epsilon$, replaced by an ``erasure state" $|2\rangle$ that is orthogonal to both $|0\rangle$ and $|1\rangle$.  This erases the input qubit and informs the receiver that it has been erased.  This channel has quantum capacity \cite{Bennett:1997mm} $Q=1-2\epsilon$ (when $\epsilon<1/2$); and when $\epsilon\geq1/2$, forward quantum communication is impossible ($Q=0$).
The fact that $Q$ strictly vanishes when $\epsilon\geq1/2$ follows \cite{Bennett:1997mm} from the ``no cloning" theorem in quantum mechanics \cite{Wootters:1982zz, Dieks:1982dj, Peres}: if Alice randomly sent half her qubits to Bob, and half to Charlie, each would experience a quantum erasure channel with $\epsilon=1/2$, and if these channels each had $Q>0$, Alice could use them to clone an arbitrary quantum message.

Our interstellar channel is an example of an erasure channel, where photons are erased in 3 ways: 

i) First, they may be erased due to ``extinction" (absorption or scattering) as they travel through the interstellar medium (ISM), from sender to receiver.  A photon of wavelength $\lambda$ traveling from a source at distance $L$ has probability $<1/2$
of extinction if $L<({\rm ln}\,2)/(n_{H}\sigma_{\lambda})$, where $n_{H}\approx 1.146 cm^{-3}$ is the typical density of Hydrogen atoms in the ISM, and $\sigma_{\lambda}$ is the ISM extinction cross section per Hydrogen atom 
at wavelength $\lambda$ \cite{Ryter, Lequeux}; or, in other words, if the horizontal line corresponding to distance $L$ lies {\it above} the blue shaded region in Fig.~1 (at wavelength $\lambda$).  For example, a sender on Proxima Centauri would have to use a wavelength where the pink dashed line (labelled Proxima Centauri) lies {\it above} the blue shaded region in Fig.~1.  

ii) Second, {\it if} we plan to receive the photon using a ground-based telescope on Earth, we must also consider extinction in the Earth's atmosphere.  For a photon to have $<1/2$ probability of being erased in {\it this} way, its wavelength must {\it also} avoid the gray bands in Fig.~1.  (On the other hand, if our receiving telescope is in space, or on the Moon, we can ignore the gray bands in Fig.~1.)

\begin{figure}
  \begin{center}
    \includegraphics[width=2.0in]{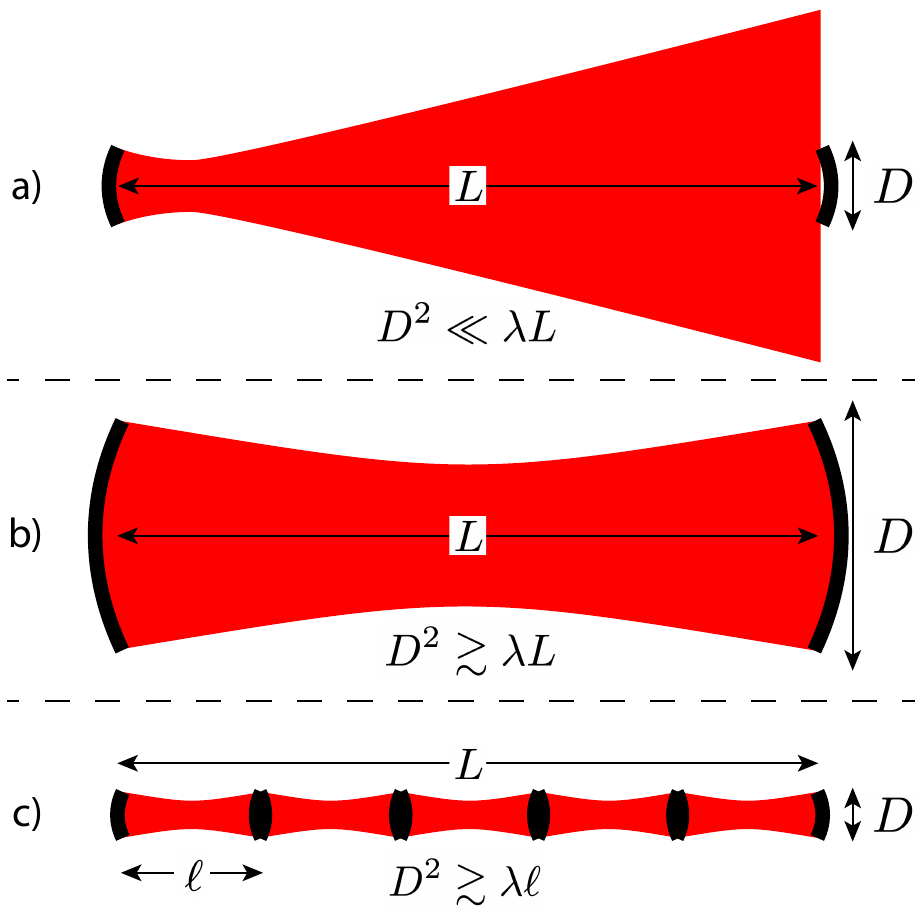}
  \end{center}
  \caption{Three interstellar channels: a) telescopes too small ($Q=0$); b) telescopes sufficiently large ($Q>0$); c) many smaller telescopes as relays ($Q>0$).}
\label{WindowFig}
\end{figure}

iii) Third, due to spreading of the photon beam as it travels from sender to receiver, the receiving telescope will only intersect a fraction of the beam (Fig.~2a), so that the remaining photons are lost.  For a photon to have $<1/2$ probability of being erased in this way requires extremely large diffraction-limited telescopes both at the transmitting end (to send a sufficiently narrow beam), and the receiving end (to encompass the beam), see Fig.~2b.  In particular, it requires (see Appendix A)
\begin{eqnarray}
  \label{D_bound}
    D&>&\left(\frac{1}{\pi}{\rm ln}\frac{2}{3-2\sqrt{2}}\right)^{1/2}(\lambda L)^{1/2} \nonumber\\
    &=&\left(\frac{\lambda}{300~{\rm nm}}\right)^{1/2}\left(\frac{L}{1~{\rm pc}}\right)^{1/2} 85.1~{\rm km}.
\end{eqnarray}
where $D=\sqrt{D_1 D_2}$ is the geometric mean of the two telescope's diameters.  Thus, $D$ is the minimum diameter for telescopes in a symmetric channel (with $D_1=D_2$), and the characteristic diameter for telescopes in an interstellar quantum communication network (with $L\sim1~{\rm parsec}$, the typical interstellar distance in the Milky Way).

This third erasure constraint is the hardest to satisfy!  Whereas {\it classical} communication ($C>0$) can take place even if the receiver only receives a tiny fraction of the photons emitted by the sender, forward {\it quantum} communication ($Q>0$) requires large enough telescopes that the sender can put the {\it majority} of their photons into the receiver's telescope (Fig.~2b)!  Even in the best case, taking the nearest star (Proxima Centauri, $L=1.30~{\rm parsec}$) and the shortest wavelength available from the ground ($\lambda=320nm$, see Fig.~1), this implies $D>100~{\rm km}$!

Is such a huge telescope even possible?  It is futuristic by current standards: the largest telescope under construction (ELT) has $D\approx 40m$.  Note that a $D\sim100~{\rm km}$ optical telescope need not be a single giant mirror: it could be a single dish with a segmented mirror (consisting of many smaller pieces, as in the largest existing telescopes, or the ELT); or a tightly-hexagonally-packed array of smaller dishes (which could cover a fraction $\pi/\sqrt{12}=0.9069$ of the area of a single dish), coherently combined, as in optical interferometry.  (Optical interferometry has so far been demonstrated over a distance of $500~{\rm m}$, while the Starshot proposal \cite{Lubin2016} has an array of coherently-combined optical telescopes with total diameter of a few ${\rm km}$.)  Given that quantum teleportation using photons has already been demonstrated over $\sim100~{\rm km}$ baselines (at sea level), and over $\sim 1000~{\rm km}$ baselines (from the Earth to a satellite), it may be that the main obstruction to building a coherent dense array of optical telescopes over $100~{\rm km}$ distances would ultimately be one of cost, rather than one of principle.  (Indeed, given that creating, manipulating and storing quantum states is currently a subject of extremely active research and ongoing progress, it is worth noting that future quantum repeaters \cite{Gottesman:2011eg} and quantum memories \cite{Bland-Hawthorn} might allow optical interferometry even over much longer baselines.) 

Using shorter wavelengths ($\lambda\ll 300~{\rm nm}$) would allow smaller $D$, but the telescopes (necessarily above the atmosphere --- e.g. on the Moon, or at Earth's L2 Lagrange point) seem even more futuristic.  For example, even if a sender on Proxima Centauri could quantumly communicate using $10~{\rm keV}$ ($\lambda\sim 10^{-10}{\rm m}$) x-rays or $1~{\rm MeV}$ ($\lambda\sim 10^{-12}{\rm m}$) gamma-rays (perhaps using nuclei as a lasing medium), this would mean a receiving telescope able to coherently catch $\gtrsim1/2$ of such photons over characteristic diameter $D\sim 2~{\rm km}$ or $D\sim 200~{\rm m}$, respectively!  

And longer wavelengths ($\lambda\gg 300~{\rm nm}$) would require even larger $D$: {\it e.g.}\ communication from Proxima Centauri with $\lambda=3~{\rm mm}$ microwave photons requires $D\sim 10^{4}{\rm km}$, comparable to the diameter of the Earth! 

{\bf Depolarizing channel.}   Sometimes, instead of receiving the sender's transmitted qubit, our telescope will receive an extraneous (astrophysical background) photon, and if the probability of this is too high, quantum communication also becomes impossible. This is described by another idealized quantum channel: the depolarizing channel, in which each input qubit $|\psi\rangle=\alpha|0\rangle+\beta|1\rangle$ is, with probability $\epsilon$, replaced by a qubit in a random state, 
without informing the receiver which qubits have been randomized.  If $\epsilon>1/3$, then $Q$ vanishes ({\it i.e.}\ no forward quantum communication); and if $\epsilon>2/3$, both $Q$ {\it and} $Q_{2}$ vanish ({\it i.e.}\ no quantum communication, even assisted by two-way classical communication)  \cite{Bennett:1996gf, Bennett:1995ra, Ekert:1996pg, Knill:1996ny}.  

If we want the randomization probability $\epsilon$ to be less than the threshhold $\epsilon_c$, then it follows from the uncertainty principle (see Appendix B) that we must restrict ourselves to wavelengths $\lambda$ such that 
\begin{equation}
  \label{I_nu_bound}
  I_{\nu}<\frac{\epsilon_{c}}{1-\epsilon_{c}}\frac{128\pi^{2}\hbar c}{\lambda^{3}}
\end{equation}
where $I_{\nu}$ (with units of ${\rm ergs}\;{\rm s}^{-1}{\rm cm}^{-2}{\rm Hz}^{-1}{\rm ster}^{-1}$) is the specific intensity of the diffuse astrophysical background at frequency $\nu=c/\lambda$.  As seen {\it e.g.} from Fig.~2.2 in \cite{Lequeux}, this constraint is easily satisfied for short wavelengths but is eventually violated at sufficiently long wavelengths by the cosmic background radiation (with temperature $T_{{\rm CMB}}=2.726~{\rm K}$), so that (\ref{I_nu_bound}) becomes (see Appendix B)
\begin{equation}
  \label{lambda_bound}
  \lambda<\frac{64\pi^{2}\epsilon_c}{1-\epsilon_c}\frac{\hbar c}{kT_{{\rm CMB}}}
  =\left\{\begin{array}{ll} 26.5~{\rm cm} & ({\rm for}\;Q>0) \\ 106\;\;{\rm cm} & ({\rm for}\;Q_{2}>0) \end{array}\right.
\end{equation}

\section{Discussion}

{\bf Motivations.}  To make our discussion more concrete, it may be helpful to give four examples to illustrate possible motivations for interstellar quantum communication:  

i) One may wish to send a complex quantum state ({\it e.g.}\ the final or intermediate state of a complex quantum computation), either directly, or via quantum teleportation \cite{Bennett:1992tv}.  Note that transmitting such an $N$-qubit state classically would mean sending $2^N$ complex numbers, which quickly becomes impossible as $N$ grows: {\it e.g.}\ for $N>265$ qubits, $2^N$ is larger than Eddington's number (the number of protons in the observable universe). 

ii) Astronomically long baseline interferometry (ALBI): As pointed out in \cite{Gottesman:2011eg}, quantum repeaters could be used to coherently interfere optical telescopes separated by the Earth's diameter $D_E$, to achieve the effective angular resolution $\delta\theta=\lambda/D_E$.  By the same token, an interstellar quantum communication channel would make it possible to interfere telescopes operating at wavelength $\lambda$, and separated by the astronomical distance $L$, thereby effectively producing a telescope with the mind-boggling angular resolution $\delta\theta=\lambda/L$.

iii) Quantum error correction: A quantum error correcting code (QECC) is a clever way of protecting a delicate quantum state from destruction by embedding it in a carefully-chosen subspace ${\cal C}$ (the code space) of a larger Hilbert space ${\cal H}$ which, in turn, may be decomposed as a tensor product ${\cal H}=\otimes_{i}{\cal H}_{i}$.  In particular, let $\rho=|\psi\rangle\langle\psi|$ be a pure state in ${\cal C}$, and let $\rho_{i}$ be the corresponding reduced density matrix in ${\cal H}_{i}$: it is a fundamental fact that ${\cal C}$ is a QECC capable of correcting arbitrary errors or erasures in ${\cal H}_{i}$ iff $\rho_{i}$ is independent of the code state $|\psi\rangle\in{\cal C}$.  Now, if each subspace ${\cal H}_{i}$ is distributed to a different solar system, we have a code with an astronomically large code distance ({\it i.e.}\ in which quantum information is protected against the erasure of the portion of the state residing in any one solar system).

iv) Quantum cryptography \cite{Bennett:2014rmv, Ekert:1991zz} allows communication whose security is guarunteed by quantum mechanics.  

{\bf Smaller telescopes?}  Is there any escape from the previous section's conclusion that an interstellar channel with $Q>0$ requires enormous telescopes?  Two loopholes are worth discussing:

i) With smaller telescopes, although forward quantum communication is impossible ($Q=0$), quantum communication assisted by two-way classical communication is possible ($Q_2>0$).  For example, imagine the sender transmits a stream of optical photons, equally spaced in time ({\it e.g.}~one per $\mu s$), each of which is a member of an EPR pair (whose other member is retained by the sender); and our telescope is smaller than the bound (\ref{D_bound}), so we randomly receive only a tiny fraction ($\ll1/2$) of these incoming photons.  If the sender doesn't know which photons we have received, they cannot use their EPR pairs to teleport their quantum states to us ($Q=0$); but if we send them a list specifying which photons we {\it did} receive (by specifying their arrival times), they {\it can} then use the corresponding subset of EPR pairs for teleportation ($Q_2>0$).  Note that, whereas forward communication (measured by $Q$) is instantaneous in the information's rest frame, communication assisted by two-way classical communication (measured by $Q_2$) involves an extra delay of at least $2L/c$ ({\it e.g.}\ at least 8 years for Proxima Centauri), and requires us to have quantum harddrives capable of storing the received photons for this duration.  Of course, this may represent an unacceptable slowdown, and would make certain tasks impossible in principle ({\it e.g.}\ if Alice wants to send two states to Bob and Charlie respectively, and have them process those states immediately, so that their spacelike separation guarantees their causal independence). 

ii) Alternatively, instead of transmission directly from sender to receiver (as in Fig.~2b), one could imagine a sequence of relays (converging lenses or quantum repeaters \cite{repeaters}) to capture and refocus the beam at $n-1$ point along its path (as in Fig.~2c).  Then, in order to achieve non-zero quantum capacity $Q>0$, the diameter of each optical element would only need to satisfy the weaker bound $D\gtrsim\sqrt{\lambda L/n}$.  In this scheme, in order to use wavelength $\lambda$ and optical elements of diameter $D$, the separation between the relay stations would have to be
\begin{equation}
  \frac{L}{n}\sim \left(\frac{D}{100~{\rm m}}\right)^{2}\left(\frac{300~{\rm nm}}{\lambda}\right)\times 3\times10^{10}~{\rm m}.
\end{equation}
So {\it e.g.}\ with $\lambda\gtrsim300~{\rm nm}$ and $D\lesssim100~{\rm m}$, the relays would need to be separated by $\lesssim0.1~{\rm au}$, with many already in our Solar System.  (Could they be detected?) Of course, placing/maintaining these repeaters in their precise locations might be too difficult/expensive in practice.

{\bf The Fermi paradox.}  Given that our universe is statistically homogeneous \cite{Maartens:2011yx, Bennett:2010jb, Planck:2018vyg, Planck:2019evm}, filled with very many galaxy clusters like our own, each containing very many galaxies rather like our own, each containing very many stars like our own, many of which have a retinue of planets, it is tempting to guess that it also contains many other occurences of life.  Fermi famously wondered why we have not yet seen any sign of them?  Many possible answers have been put forward \cite{Webb, Forgan}.  Here we point out a new answer suggested by the preceding considerations:

Suppose that the sender wishes to communicate quantum rather than classically.  (We have mentioned several possible motivations for this, and there will certainly be many more: as mentioned above, classical communication is just a part of the larger topic of quantum communication, whose limits and applications are still only partially understood.)   Two simple conclusions then follow from the requirement of non-zero quantum capacity $Q>0$.  (i) First, we have seen that the sender must place nearly all (at least $1/2$) of their photons into our receiving telescope, which implies that the signal must be so highly directed that only the intended receiving telescope can hope to detect any sign of the communication.  This is in sharp contrast to classical communication, where one can broadcast photons indiscriminantly into space, and an observer in any direction who detects a small fraction of those photons can still receive the message.  (ii) Second, we have seen that (setting aside the loopholes mentioned above) the sending and receiving telescopes must be extremely large, satisfying the inequality in Eq.~(\ref{D_bound}); but this same inequality implies that, if the sender has a large enough telescope to communicate quantumly with us, they necessarily also have enough angular resolution to {\it see} that we do {\it not} yet have a sufficiently large receiving telescope \footnote{(and they have the resolution needed to determine the position of our telescope, or where it will be 4 years hence, if they are communicating from Proxima Centauri)}, so it would make no sense to send any quantum communications to us until we had built one.  Thus, the assumption that interstellar communication is quantum appears sufficient to explain the Fermi paradox.

{\bf Acknowledgements.}  I am grateful to Daniel Gottesman and Avi Loeb for very helpful discussions.

\section{Appendix A}

Here we derive Eq.~(\ref{D_bound}).  Actually, we give a sequence of three increasingly precise derivations, to make it clear how various descriptions that the reader may be familiar with relate to one another.

1) For starters, let us make a rough estimate (ignoring factors of order unity) of the telescope sizes needed for one telescope to catch an order-one fraction of the photons emitted by the other telescope.  The transmitting telescope of diameter $D_1$ can aim an electromagnetic beam of wavelength $\lambda$ with, at best, the diffraction-limited angular uncertainty $\Delta\theta\sim\lambda/D_1$.  Thus, after traveling a distance $L$ (in the $z$ direction), the beam will have spread (in the $xy$ plane perpendicular to the $z$ axis) to a characteristic width $L\Delta\theta\sim\lambda L/D_1$.  This width must be $\lesssim D_{2}$ (the diameter of the receiving telescope) or, equivalently,
\begin{equation}
  D_1 D_2 \gtrsim \lambda L  
\end{equation}
in order for the receiving telescope to catch an order-one fraction of the photons. 

2) To phrase things more precisely (but still classically), consider the usual model for an ideal laser beam: an axisymmetric beam of electromagnetic radiation, freely propagating along the $z$ axis, with a Gaussian profile in the transverse $xy$ plane.  The beam is described by (see {\it e.g.}\ Eq. 8.40 in \cite{ThorneBlandford})
\begin{equation}
  \psi_{z}=\frac{\sigma_{0}}{\sigma_{z}}{\rm exp}\!\left(\frac{-\rho^{2}}{4\sigma_{z}^{2}}\right)\!
  {\rm exp}\!\left[i\left(\frac{k\rho^{2}}{2R_{z}}\!+\!kz\!-\!{\rm arctan}\frac{z}{z_{0}}\right)\right]\!
\end{equation}
where $\rho=(x^{2}+y^{2})^{1/2}$ is the distance from the $z$ axis, $k=2\pi/\lambda$ is the wavenumber of the beam, and we have defined the quantities $z_{0}=2k\sigma_{0}^{2}$,  $\sigma_{z}=\sigma_{0}(1+z^{2}/z_{0}^{2})^{1/2}$, and $R_{z}=a(1+z_{0}^{2}/z^{2})$.  At fixed $z$, the beam's energy flux distribution is $\propto|\psi_z|^{2}=(\sigma_0/\sigma_z)^2 {\rm exp}[-\frac{1}{2}\frac{\rho^{2}}{\sigma_z^2}]$, so as the beam propagates along the $z$-axis, it retains its gaussian profile, with beam radius $\sigma_0$ at its waist ($z=0$), and beam radius $\sigma_z$ at a general $z$.  Now consider the product $\sigma_z \sigma_{z\pm L}$ between the beam radii at two points along the $z$ axis separated by distance $L$.  The product $\sigma_z\sigma_{z\pm L}$ is minimized when $\partial(\sigma_z \sigma_{z\pm L})/\partial\sigma_0=0$, which yields
\begin{equation}
  \label{sigma_1_sigma_2}
  \sigma_{z}\sigma_{z\pm L}\geq \lambda L/4\pi,
\end{equation}
in agreement with our previous rough estimate.

3) Now let us phrase things quantum mechanically: Suppose a photon is emitted in the $z$-direction with $z$-momentum $p_{z}=\hbar k=2\pi \hbar/\lambda$.  If its initial position uncertainty (along some transverse direction) is $\Delta x_1$, then (by the uncertainty principle) its corresponding transverse momentum uncertainty is $\Delta p_{\perp}\geq \frac{1}{2}\hbar/\Delta x_1$, so the angular uncertainty in its direction is $\Delta\theta=\Delta p_{\perp}/p_{z}\geq\lambda/(4\pi\Delta x_1)$ and, after propagating a distance $L$ along the $z$-axis, its transverse position uncertainty has grown to $\Delta x_2= L\Delta\theta\geq\lambda L/(4\pi\Delta x_1)$ or, equivalently,
\begin{equation}
  \label{Delta_x1_Delta_x2}
  \Delta x_1 \Delta x_2 \geq \lambda L/4\pi,
\end{equation}
which again agrees with the previous result (\ref{sigma_1_sigma_2}).  

Next take the photon's transverse wavefunction to be described by an axisymmetric gaussian $\psi(\rho)$ which saturates the position-momentum uncertainty bound, so that Eq.~(\ref{Delta_x1_Delta_x2}) becomes an equality.  The photon has transverse probability distribution $|\psi_{1,2}|^{2}=\frac{1}{2\pi\Delta x_{1,2}^2}
{\rm exp}(-\frac{1}{2}\frac{\rho_{1,2}^{2}}{\Delta x_{1,2}^2})$, where the subscripts $1$ and $2$ apply at the two ends of the $z$-axis, respectively; so the joint-probability 
that the wavefunction will overlap with both the initial and final telescopes (of diameter $D_1$ and $D_2$, respectively) is
\begin{equation}
  \left[1-{\rm exp}\left(\frac{-D_{1}^{2}}{8\Delta x_1^{2}}\right)\right]\left[1-{\rm exp}\left(\frac{-D_{2}^{2}}{8\Delta x_2^{2}}\right)\right],
\end{equation}
and the requirement that this is $\geq 1/2$ implies Eq.~(\ref{D_bound}).

\section{Appendix B}

Here we derive Eqs.~(\ref{I_nu_bound}) and (\ref{lambda_bound}).  

Consider a transmitted qubit photon of wavelength $\lambda$ that arrives with uncertainty $\Delta t$ in its arrival time, $\Delta E$ in its energy, and hence $\Delta\nu=(\Delta E)/(2\pi\hbar)$ in its frequency.  Due to the energy-time uncertainty relation, we have $\Delta t \Delta E\geq \hbar/2$, and hence $\Delta t \Delta\nu\geq 1/4\pi$.   The number of astrophysical background photons arriving within the time interval $\Delta t$ and frequency range $\Delta\nu$ is
\begin{equation}
  \label{p1}
  N=\frac{I_{\nu}\cdot \Delta t \cdot \Delta \nu \cdot A \cdot\Delta\Omega}{2\pi\hbar\nu}
\end{equation}
where $I_{\nu}$ is the specific intensity of the astrophysical background (energy per time per frequency per area per solid angle) at frequency $\nu=c/\lambda$, $A$ is the receiving telescope's area, $\Delta\Omega$ is its angular resolution (in solid angle), and $2\pi\hbar\nu$ is the energy per photon.  If the receiving telescope has diameter $\Delta x$, and hence area $A=\pi(\Delta x)^2$, a received photon has transverse position uncertainty $\Delta x$, hence transverse momentum uncertainty $\Delta p \geq \hbar/(2\Delta x)$, hence angular uncertainty $\Delta\theta\geq(\Delta p)/p=\lambda/(4\pi \Delta x)$, and hence angular resolution $\Delta\Omega=\pi(\Delta\theta)^2\geq(1/16\pi)(\lambda/\Delta x)^2$.  Thus, (\ref{p1}) becomes
\begin{equation}
  N\geq \frac{I_{\nu}\lambda^{3}}{128\pi^{2}\hbar c}
\end{equation}
On the other hand, if we write $N$ (the expected number of random photons per signal photon) as $\epsilon/(1-\epsilon)$, and solve for $I_{\nu}$, we obtain Eq.~(\ref{I_nu_bound}), the first desired result.

From Fig.~2.2 in \cite{Lequeux}, $I_{\nu}$ satisfies the bound (\ref{I_nu_bound}) for wavelengths $\lambda<\lambda_{CMB}^{}$ (where $\lambda_{CMB}$ is the peak of the cosmic microwave background), but eventually violates it for wavelengths $\lambda\gg\lambda_{CMB}^{}$.  So, substituting 
\begin{eqnarray}
  I_{\nu}^{{\rm CMB}}&=&\frac{4\pi \hbar c}{\lambda^{3}}\left[{\rm exp}\left(\frac{2\pi \hbar\nu}{kT_{{\rm CMB}}}\right)-1\right]^{-1} \nonumber\\
  &\approx&\frac{2kT_{{\rm CMB}}}{\lambda^{2}}\qquad(\lambda\gg\lambda_{{\rm CMB}}^{})
\end{eqnarray}
into Eq.~(\ref{I_nu_bound}), and solving for $\lambda$, we obtain Eq.~(\ref{lambda_bound}), the other desired result.  

The specific intensity $I_{\nu}^{sender}$ of the sender's photons must exceed $\frac{1-\epsilon_c}{\epsilon_c}I_{\nu}$ in the pixel corresponding to their direction on the sky (which sets a minimum rate at which they must transmit photons); but, since a channel satisfying (\ref{D_bound}) has the angular resolution to distinguish the sender from their star, this can always easily satisfied at wavelengths satisfying (\ref{lambda_bound}).

\end{document}